\documentclass{article}

\usepackage[utf8]{inputenc} 
\usepackage[T1]{fontenc} 
\usepackage{geometry} 
\usepackage{titlesec} 
\usepackage{fancyhdr} 
\usepackage{graphicx}
\usepackage{natbib}
\usepackage{url}
\usepackage{setspace}
\usepackage{xcolor}
\usepackage{tikz}
\usepackage{tikzsymbols}

\title{Speculations on Uncertainty and Humane Algorithms}
\author{Nicholas Gray}
\date{University of Liverpool, UK\\ngg@liverpool.ac.uk}

\begin{document}

\maketitle

\section*{Abstract}
The appreciation and utilisation of risk and uncertainty can play a key role in helping to solve some of the many ethical issues that are posed by AI. Understanding the uncertainties can allow algorithms to make better decisions by providing interrogatable avenues to check the correctness of outputs. Allowing algorithms to deal with variability and ambiguity with their inputs means they do not need to force people into uncomfortable classifications. Provenance enables algorithms to know what they know preventing possible harms. Additionally, uncertainty about provenance highlights the trustworthiness of algorithms. It is essential to compute with what we know rather than make assumptions that may be unjustified or untenable. This paper provides a perspective on the need for the importance of risk and uncertainty in the development of ethical AI, especially in high-risk scenarios. It argues that the handling of uncertainty, especially epistemic uncertainty, is critical to ensuring that algorithms do not cause harm and are trustworthy and ensure that the decisions that they make are humane.

\textbf{Keywords}: \textit{AI Ethics, Uncertainty in AI, Trustworthy AI}

\section{Introduction}
Increasingly inhumane algorithms adjudicate more and more of our daily lives. 
Algorithms recommend our entertainment, decide who to give loans to, mediate social and business interactions and decide who should go to prison.
This increasing algorithmisation has numerous risks ranging from irritation and confusion to injustice and catastrophe.

Despite these risks, society is fast moving towards ``algorithm appreciation'' \citep{Logg2019} with many going out of their way to do things using automated systems. There is an increasing appetite to exploit big data and machine learning tools in almost every field, including risk analysis \citep{Nateghi2021}. There is a risk of treating algorithms as epistemic superiors that can perform tasks faster and far better than other humans. With some trusting in AI over their own decision-making \citep{Logg2019}, this depends on the context of the decision, with human decision-making still preferred in medicine, economics and education \citep{Kaufmann2021, Kerstan2023, Schwienbacher2020, Hou2021}.

There is an increasing narrative about the existential danger presented by AI \citep{Bender2021, Carlsmith2022, Hendrycks2022, Ordonez2023, Weidinger2021}, and the speed at which machine learning technology is advancing has made some call for the development of the technology to be paused \citep{FutureofLifeInstitute2023}. Ultimately, no tech-Cnut will be able to hold back the tide and instead the focus must move on to assessing the risks presented by algorithms.\footnote{ King Cnut, (sometimes spelt Canute), was an 11th-century English monarch who apocryphally attempted to command the tide to stop and got his feet wet.}
Instead, there is a need to ensure that AI systems are designed to inflict minimal harm, with the focus on who benefits from increasing AI involvement \citep{Goldberg2024}, they need to be designed to be obliging and compassionate towards their end users, they need to be more humane.\footnote{ The idea that algorithms need to be humane is inspired by \citet{Kahan1992, Kahan2011}.}

Among the many published ethical principles and guidelines for AI, there is a consensus around five key themes: non-maleficence, transparency (including explainability), justice and fairness, responsibility and accountability, and privacy \citep{Jobin2019}.
Much has already been written about solutions to all of these problems. Often overlooked, however, is the importance of risk and uncertainty as an important tool for solving many of these problems. For example, deliberately introducing uncertainty into datasets can help protect individuals from reidentification \citep{Kreinovich2015, Longpré2007, Longpré2013, Longpré2017}. 
For this discussion, we will consider how uncertainty is fundamental in answering two key questions that are critical to addressing any\footnote{ Perhaps with the exception of privacy protection.} of the ethical issues \citeauthor{Jobin2019} raised:
\begin{enumerate}
 \item Is the output of the algorithm correct? In both empirical and moral sense, and
 \item How much control can be ceded to the algorithm? 
\end{enumerate}
First we need to consider the significance of the numbers that are used within algorithms.

\section{The Numbers of the Future}\label{sec:numbers}
Shortly after the invention of Charles Baggage's difference engine, which can be considered a very early ancestor of today's computers, he was asked:
\textit{``Prey, Mr. Babbage, if you put into the machine the wrong numbers will the right answers come out?''} \citep{Babbage1864}.
He was ``not able rightly to apprehend the kind of confusion of ideas that could provoke such a question.''
Whilst he could not have possibly imagined the distant descendant of the machines that he created would be so ubiquitous within today's society, he would still not be able rightly to apprehend that people continue putting the wrong numbers into the machine and expecting the right answers to come out.

This incorrectness is not just caused by the ``garbage in, garbage out'' concept, but the fact that numbers are not some abstract philosophical or numerological notion of \textit{fiveness} but instead are answers to questions such as ``how many?'' or ``how long?''; or quantitatively answers to ``where?'' or ``when?''\footnote{ This does exclude numbers that are used for purely nominal reasons (e.g. room 101). Although, the origin of these is often in counting.} These numbers are rarely precise values; they have units, they have provenance, they have uncertainty.
Yet, algorithms often strip this away by requiring unitless, contextless, precise inputs to perform calculations.

For example, when a nurse needs to calculate the quantity of a drug a patient can safely consume, it is often trivial arithmetic. It is also trivially easy for there to have been a calculation mistake. Simply misplacing a decimal point can be fatal. Simple unit errors harm thousands of children each year \citep{Kuehn2015}.
Analysis of these incidents often overlooks the algorithm factor--it is unlikely to have made a mistake--and instead focuses on human factors \citep{Thimbleby2010a}. Calculators can be designed to detect and block such mistakes by knowing what drug it is and what units are being used \citep{Thimbleby2010}. By being aware of the context of the numbers, algorithms can ``sense check'' results and block incorrect calculations thus reducing harm. This can help make algorithms more humane by improving their trustworthiness.

\subsection{Uncertainty}
There are two types of uncertainty: aleatory uncertainty arises from the natural variability of systems and epistemic uncertainty due to a lack of knowledge of a system. Epistemic uncertainty might arise from measurement errors, missing data, censoring of data or ambiguity. Aleatory uncertainty is naturally characterised by probability distributions. Epistemic uncertainty is often imagined to be covered using Bayesian methods, or it is often ignored or assumed away when training machine learning models, to present a single middle-of-the-road model \citep[Chapter~5]{THESIS}. This means that the discussion of uncertainty in AI has been focused on probabilistic methods.\footnote{ See \citet[Appendix C]{Russell2019} as an example.} Nevertheless, this is debatable at best because it is challenging to represent the lack of information using a single probability, especially for risk analysis \citep{Aven2010, Hullermeier2021, PBA-paper}.

For example, risk-averse algorithms are less likely to approve loans to individuals from minority ethnic groups for which there is a lack of data, something \citet{Goodman2016} called \textit{uncertainty bias}. This bias is artificial and can occur irrespective of any bias within the dataset. It comes from the conflation of the two types of uncertainty: aleatory uncertainty, meaning that for individuals with similar demographic and financial characteristics, some will default and some will not, and epistemic uncertainty caused by the lack of information about the minority groups. The algorithm is not able to express the fact that it does not know how likely it is that an individual will default and must therefore reject.

Enabling users to express uncertainty within inputs may also help to guard against hermeneutic injustice: where a person is disadvantaged because they are unable to make their experiences intelligible \citep{Fricker2007, Walmsley2020}. For example, a person may not communicate their symptoms accurately to a medical decision support algorithm. The algorithm might want to know how long the patient has had a cough but only accepts precise answers such as 6 weeks, whereas the patient might only provide an estimate such as ``at least three weeks but no more than two months'', or simply not know the answer to the question. 

\subsection{Provenance}
Provenance records the origin and ownership history of an algorithm or dataset. For example, knowing the provenance of a medical decision support tool would enable answering the following:
Why a particular machine learning algorithm has been used? Where did the training data come from? What was the data cleaning process? Who made what assumptions and why? Who owns the model and can modify it? Assuring the trustworthiness of such algorithms requires knowledge of their provenance, and any uncertainty may suggest otherwise.

There is a multiverse of different models that could have been fitted from the same data \citep{Steegen2016}. This excludes the different multiverses that could have been created through the selection of machine learning algorithms, hyperparameters, and the data cleaning process. \citet{Riley2023a} have shown that fitting models with different representative samples from the same population can lead to models that produce significantly different results. Models can be highly unstable--implying large uncertainty in models--especially with low sample sizes. This uncertainty means that one model within the multiverse is not necessarily more correct than another and different models may make different decisions \citep{Riley2023b}.
Whilst bootstrapping can be used \citep{Riley2023a}, knowing the provenance of the data is the best way to check for this instability and assess the correctness of the model.

Knowing the provenance can also have other benefits.
There is more disinformation than information on social media about COVID-19 \citep{edwards-newcastle}, with much of it originating from spurious claims made by only twelve individuals \citep{CCDH-fail, CCDH-12}. 
Being aware of the origins of the claims may allow algorithms to detect dangerous falsehoods when recommending content \citep{Baeth2018}. 

Generative AI systems are likely to exacerbate these problems, making the question of prevalence even more critical. 
For instance, after the 2022 Russian invasion of Ukraine, deep-fake images were being circulated propagandically \citep{Farid2022}. The ability of generative AI systems to flood the internet with misinformation risks undermining trust in legitimate sources \citep{Epstein2023}.
There is also the risk of generative AI systems `hallucinating'--returning unintended text that is often unfactual or incorrect in some other way \citep{Maynez2020, Ji2023}. There are numerous potential harms, from making mistakes falsifying scientific references \citep{Alkaissi2023, Emsley2023}, insider trading \citep{Wain2023}, or encouraging high treason \citep{Zaccaro2023}. 

Provenance is a potential solution to many problems, helping algorithms to highlight their trustworthiness \citep{Ferrara2023}, but uncertainty can also play its part \citep{Xiao2022}, since many hallucinations may be caused by algorithms inventing facts when faced with uncertainty \citep{Lightman2023}.
Uncertainty in provenance means that it is likely the case that one is not able to trust that the outputs are correct or trustworthy. If a large language model can express what they don't know then, as opposed to generating nonsense--but not-nonsensical--prose, then any hallucinogenic harms may be avoided.

There are lots of potential benefits of algorithms being able to uncover unknown-knowns, information that exists within datasets that could only be uncovered by big data or machine learning methodologies. However, it is only by appreciating the known-unknowns that algorithms can be more humane.

\section{Uncertainty Ex Machina} \label{sec:ex-machina}

\subsection{Is the algorithm correct?} 
There are numerous different approaches to answering this question with the most obvious being to use statistics to measure the empirical accuracy of the output. There are many such statistics to choose from and they are often context dependent. For classifiers, two of the most popular methods are Receiver Operating Characteristic (ROC) curves along with the area under them (AUC/C-statistic), and statistics derived from confusion matrices, such as accuracy, sensitivity/specificity, precision/recall, F1 score, etc.

It is not the case that decision-making is analogous to classification, especially in high-risk situations. Decisions are often more nuanced than a simple yes/no and different errors have different costs. 
The development of these tools needs to not be seen as a tournament of algorithms competing to be the most empirically correct. The ability of large language models to pass medical or legal exams \citep{BommaritoII2022, Jung2023, Singhal2023}, says nothing of their ability to perform these talks in the real world. 
Imagine an algorithm is used to predict the likelihood of somebody attempting suicide. Numerous false positives may be annoying but a single false negative is likely to be catastrophic. 
It needs to be acknowledged that a patient presentation is a unique instance and that their future will be influenced by a unique set of countless environmental, physiological and psychological factors in constant interplay, ergo using a single dataset to inform high-risk decision making is flawed \citep{Nathan2021, Nathan2024}.

Assessing the empirical accuracy alone is not sufficient when considering the correctness of algorithms. Defining moral correctness is harder as it depends on what is deemed sociologically acceptable although bad apples are often easy to spot and are widely reported. For example, Apple and Goldman Sachs were accused of systematically granting men higher credit card limits than women \citep{apple-card-sexist}.
Much of the AI literature on fairness is focused on defining metrics for bias that algorithms can assess so that they can be designed to achieve an appropriate morality level \citep{Corbett-Davies2023, Jacobs2021, Jui2024, Milkolajczyk2023, Pagano2022}.
There is unlikely to ever be a precise algorithm for fairness, there will always be conflicting and ambiguous definitions of fairness, and as such being able to correctly handle uncertainty will be fundamental to counteract the risk of unfairness.

Explainability is widely considered another important aspect to consider the correctness of algorithms \citep{Doshi-Velez2017, Gilpin2018}. This often takes the form of a second post-hoc model being used to explain the first black box model \citep{Rudin2019, Angelov2021}. The idea is: if the black box answers the \textit{what?/how much?/etc}, then the explainability part answers the \textit{why?/how?}

Some argue that explainability is not the remedy that is needed to protect against malicious algorithms \citep{Edwards2017}. At the same time, explainability may not be possible, and there are situations where a complete understanding of the algorithm is neither desired nor required \citep{Walmsley2020}. 
It has even been argued that in high-risk domains the use of black-box machine learning models, even if they come with post-hoc explainability, should be entirely avoided. Instead, effort should be focused on ensuring models are naturally interpretable (i.e. they produce logical rules or rely on classical statistical methodologies) \citep{Rudin2019, Joyce2023}.

\begin{figure}[t]
 \centering
 \includegraphics[width=0.5\textwidth]{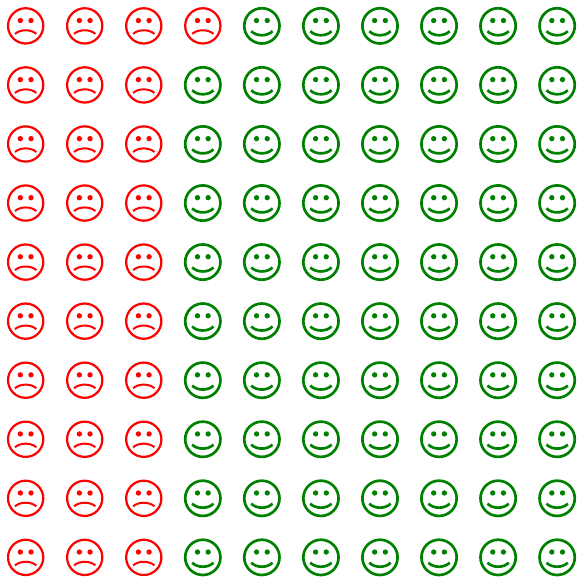}
 \caption{An example icon array with $\Pr=0.31$, 31 out of 100 are at risk of sleep apnea.}
 \label{icon-array}
\end{figure}

Aside from accuracy and explainability, letting AI express epistemic uncertainty is critical to assessing its correctness and gives another intelligible avenue for appeal. 
Although many algorithms are inherently probabilistic, outputting naked probabilities as though they are the beginning and end of an algorithm's uncertainty is misguided. 
BASH-GN is a medical risk prediction tool that assesses the risk of a patient having obstructive sleep apnea \citep{Huo2023}.\footnote{ It is accessible at \url{https://c2ship.org/bash-gn-metric/}} The output of the model is a naked probability alongside a low/high-risk classification (e.g. 30.9\% - \texttt{low risk}\footnote{ This result is produced with the following inputs: female, aged 54, neck circumference 24cm, weight 90kg, height 1.56m, with high blood pressure and snoring as loud as talking. Pleasingly the algorithm allows users to select their preferred unit between metric and United States customary units.}). 
This presentation of risk as a single probability is often misunderstood \citep{Aven2010, Aven2023, Gigerenzer2005, Wegwarth2011}.
Improvements can be made by presenting information in ways people find easier to understand. Natural frequencies are a more intuitive way for people to understand probabilities, and it is easy for algorithms to return expressions such as ``31 out of 100'' instead of $\Pr=31\%$ \citep{Gigerenzer1995, Gigerenzer2011}. Icon arrays are another approach that can be used to express information from algorithms \citep{Galesic2009}, see Figure~\ref{icon-array} for an example. They show natural frequency information using pictograms that are easy to understand even for people with poor numeracy skills \citep{Zikmund-Fisher2014}. 

The use of natural frequencies or icon arrays would also help express the context. It is possible to interpret $\Pr=31\%$ in several different ways:
\begin{enumerate}
 \item 31\% of patients with similar symptoms suffer from sleep apnea.
 \item 31\% of nights the patient suffers from sleep apnea.
 \item 31\% of the time the patient is asleep, the patient has sleep apnea episodes.
\end{enumerate}
Each of these different interpretations would have a significant impact on any decisions about further testing or treatments for doctors or patients.

Expressing the algorithm uncertainty also aids trustworthiness. 
BASH-GN outputting $\Pr=31\%$ implies a level of confidence that may be unwarranted. If the algorithm outputted an interval probability, the output $\Pr=[29,33]\%$ implies that the result is stable and reliable, whereas if the output was $\Pr=[5,95]\%$, then the vacuousness of this result suggests that the algorithm should not be relied on in the decisions making process. 
Expressing this uncertainty to the user is critical. \citeauthor{Goodman2016}'s uncertainty bias is in part because the uncertainty is reduced to a single value\footnote{ In their example, the centred 95\% confidence interval is reduced to the lower bound and compared to the decision threshold.} and this the fact that there is not enough information to make a reliable decision is never expressed outside of the black box. The fact that it cannot say ``I don't know'' is what leads to the biased decision making.

Uncertainty can also be expressed using natural language expressions. These can be done using approximators (``about 5''), plausibility shields (``I think it's 5'') or strengtheners (``I am certain it's 5''). Additionally expressing the provenance of the information can highlight the level of trust one should place in its provenance (``According to the Daily Mail/Fox News, it's 6'') \citep{vanderBles2019, Amayuelas2023, Zhou2023}. All these statements express the algorithm's uncertainty in transparent ways but in general, such an approach will need to be carefully constructed, ensuring that the gist of the output is clear to the users and not obstructed with unnecessary information. 

Uncertainty plays a key part when assessing the correctness of algorithms by allowing humans to understand the results, with natural expressions of confidence.
High uncertainty gives information about how untrustworthy the output may be, which can help shield against potential harm.
Expressing the uncertainty within the output may allow for better decisions to be made and can give a clear avenue to appeal a decision made by an algorithm. It is unclear what the best approach to expressing uncertainty is, but it is clear that it is important to do so. 
The users that need the uncertainty may not be able to understand probabilities or intervals and for them, all that is needed is a simple recommendation along with an expression of confidence, ideally in an explainable manner. Other users may need more detailed information, such as the full distribution of the output, to make a decision.
The output does not matter to the algorithm, but it matters to the people who have to interact with it, especially in high-risk scenarios.

\subsection{Cede to the algorithm?}
The human-algorithm relationship needs to be adequately defined, especially if the algorithm may be accidentally maleficent.
There are numerous examples of algorithms abetting disasters or injustices, irrespective of whether the algorithm was bypassed or trusted blindly; gave-up or refused to relinquish control. The number of such incidents is increasing as more and more decisions are ceded to algorithms.

\textit{The Smiler} is a rollercoaster at Alton Towers theme park in the UK. In 2015, an accident occurred when two cars on the track crashed into each other resulting in two people needing leg amputations. The ride operators bypassing the safety algorithm was a direct cause of the tragedy. The algorithm was designed to ``err of the side of caution'' ``because [it] does not, and cannot, have eyes, the signals from the [sensors] are a proxy for reality, and may not always accurately reflect reality'' \citep[p.~17]{Flanagan2015}. This erring of the side of caution caused numerous false alarms causing the operators to believe that the algorithm was ``crying wolf'', meaning that the operators ignored the error without justification leading to disaster. 

Within criminal justice (and presumably across every field), ``algorithms provide [decision makers] with a way to do less work while not being accountable.''\citep{Berk-in-HelloWorld} and using algorithms ``shifts accountability [...] to black-box machinery that purports to be scientific, evidence-based and race-neutral'' \cite{Lum2016}. However, the fact that algorithms are `scientific' and `evidence-based' by no means ensures that they will lead to accurate, reliable, or fair decisions \citep{Barocas2019}, with the most notorious criminal justice algorithm, COMPAS used to predict recidivism, having major racial bias problems \citep{Angwin2016, Dressel2018, Larson2016}. 

Air France Flight 447 (AF447) was an Airbus A330 aircraft flying from Rio De Janeiro to Paris in 2009. 
When the pitot tubes that measure airspeed froze over, the autopilot--faced with uncertain speed information--disengaged, passing over control of the aircraft to the two pilots in the cockpit. 
Unused to flying the aircraft at 38,000 ft, the pilots made a series of errors, stalled the aircraft and caused it to crash into the ocean killing all on board \citep{BEA2012}.
Expecting the overseeing humans to `save the day' in this way can be dangerous, not least as there is the possibility that they have forgotten how to do the task in the first place, a problem that \citet{Bainbridge1983} called \textit{the irony of automation}. 

The Boeing 737-MAX had a Manoeuvring Characteristic Augmentation System (MCAS) that suffered from almost the reverse problem to that of AF447. MCAS was designed to ensure that the flight control system mimicked the behaviour of older generation aircraft by adjusting the control surfaces when it detected, using a single angle of attack sensor, the aircraft may be approaching stall \citep{Hart2019}. Two crashes occurred as the pilots of the aircraft were unable to shut MCAS off completely when sensors started receiving incorrect readings after take-off \citep{Cusumano2021, ECAA2019, KNKT2019}. Boeing implicitly decided that the algorithm would never make an error and ensured that it could never fully cede control to the human pilots.

The expression of unsureness is critical to helping prevent these problems. Trusting that an algorithm can say ``I don't know'' when it cannot make a decision can be of fundamental importance in helping to make the lives of its human supervisors easier. This in turn can help prevent incorrect results by providing a framework for the algorithm to ask for human intervention. If The Smiler could have distinguished between when it knew the track was not safe and when it did not know that the track was safe then the supervisors might have trusted it more. 
If COMPAS was able to communicate uncertainty about its decision in the ways discussed above, instead of a simple low/high risk, it may have prevented judges from blindly trusting its outputs, which may have reduced injustices. Such uncertainty communication could also help alleviate the uncertainty bias problem discussed above.
If the A330 autopilot had been able to communicate the uncertainty in the airspeed measurement to the pilots and given them contextualised information to help address it, instead of leaving them (literally) in the dark, it may not have crashed. Equally, if the 737-MAX's MCAS system had understood that its single sensor did not always accurately reflect reality, then those disasters might not have occurred. 

There is unlikely to ever be a universal algorithm that can be deployed to prevent injustices or catastrophes, the important thing is that in high-risk situations algorithms are designed to act humanely.

\section{Conclusions}
Algorithms need to be designed to work harmoniously with humans, avoiding tyrannical or harmful behaviour. To prevent injustices and catastrophes, they should produce outcomes that are equitable and fair by recognizing and accommodating user diversity, accepting varied human inputs, and being aware of the context and provenance of the data they use. Transparency is essential; algorithms should be understandable and explainable to help humans verify their outputs and ensure correctness. Trustworthiness in control requires that algorithms accept and relinquish control in ways humans find workable, incorporating fail-safe measures to prevent disasters. Additionally, they must protect privacy by securing or anonymizing personal information.

Only through properly handling uncertainty in high-risk scenarios can humane algorithms be developed.

\section*{Acknowledgements}
This paper has benefitted from discussions with many people including Daniel Joyce, Dominic Calleja, Enrique Miralles-Dolz, Adolphus Lye and Alexander Wimbush. Special thanks to Scott Ferson for proof-reading a draft of the manuscript.



\end{document}